\documentclass[twocolumn,floatfix,nofootinbib]{revtex4-1}
\pdfoutput=1

\usepackage{hyperref}
\usepackage{graphicx}
\usepackage{xspace}
\usepackage{amsmath}
\usepackage{amssymb}
\usepackage{enumitem}
\usepackage[babel]{microtype}

\usepackage[skip=6pt]{caption}

\graphicspath{{./figs/}}

\usepackage[english]{babel}
\usepackage[T1]{fontenc}
\usepackage[utf8]{inputenc}
\usepackage[scaled=1.04]{biolinum}

\usepackage{fourier}
\usepackage[scaled=0.83]{beramono}

\newcommand{\lne}{%
  \raisebox{-0.12cm}{%
    \shortstack{%
      \ensuremath{\le} \\[-0.13cm] \scalebox{1.5}{\textcolor{white}{--}}%
    }%
  }%
}

\DeclareMathOperator*{\approxless}{%
  ~\raisebox{-0.08cm}{%
    \shortstack{%
      \ensuremath{\lne} \\[-0.25cm] \scalebox{0.9}{\ensuremath{\sim}}
    }%
  }%
~
}

\usepackage{natbib}

\makeatletter
\def\NAT@spacechar{\,}
\makeatother

\hypersetup{
    pdftitle={Bayesian and frequentist approaches to resonance searches},
    pdfauthor={Andrew Fowlie},
    colorlinks=true,
    allcolors=[rgb]{0.26,0.41,0.88},
}

\let\oldcite\cite
\renewcommand{\cite}{\,\oldcite}

\newcommand{\refsec}{sec.\,\ref}
\newcommand{\refcite}{ref.\,\oldcite}
\newcommand{\Refcite}{Ref.\,\oldcite}
\newcommand{\reffig}{fig.\,\ref}
\newcommand{\refeq}{eq.\,\ref}

\newcommand{\refprior}{prior~\ref}

\newcommand{\gev}{\,\text{GeV}}
\newcommand{\tev}{\,\text{TeV}}
\newcommand{\fb}{\,\text{fb}}
\newcommand{\invfb}{/\text{fb}}
\newcommand{\lumi}{\mathcal{L}}

\newcommand{\dif}{\,\text{d}}

\newcommand{\data}{d}
\newcommand{\given}{\, \middle| \,}
\newcommand{\prob}[2]{p\left(#1 \given #2 \right)}
\newcommand{\prior}[1]{p\left(#1\right)}

\newcommand{\Prob}[2]{P\left(#1 \given #2 \right)}
\newcommand{\Prior}[1]{P\left(#1\right)}

\newcommand{\subbox}[1]{\raisebox{0.5pt}{\scalebox{0.8}{#1}}}

\DeclareMathOperator{\Gaussian}{\phi}
\newcommand{\pvalue}{\text{\textit{p-}value}\xspace}
\newcommand{\pvalues}{\textit{p-}values\xspace}
\newcommand{\zvalue}{\text{\textit{Z-}value}\xspace}
\newcommand{\zvalues}{\textit{Z-}values\xspace}
\newcommand{\CLs}{\ensuremath{\text{CL}_s}}

\newcommand{\notE}{q < q_\text{Observed}}

\newcommand{\ncumulative}{\text{$10$\,k}\xspace}
\newcommand{\npseudo}{\text{$100$\,k}\xspace}
\newcommand{\nprior}{fourteen\xspace}
\newcommand{\naltprior}{thirteen\xspace}
\newcommand{\nvague}{seven\xspace}
\newcommand{\nspecific}{seven\xspace}
\newcommand{\nlumi}{50\xspace}

\newcommand{\figwidth}{0.8\linewidth}
\newcommand{\widefigwidth}{0.7\linewidth}
\newcommand{\todash}{\text{ -- }}
\newcommand{\labelenu}[1]{\addtocounter{enumi}{-1}\refstepcounter{enumi}\label{#1}}
\begin{document}

\title{Bayesian and frequentist approaches to resonance searches}
\author{Andrew Fowlie\vspace{0.2cm}}
\affiliation{Department of Physics and Institute of Theoretical Physics, Nanjing Normal University, Nanjing, Jiangsu 210023, China}
\email{Andrew.J.Fowlie@qq.com}
% \date{\today}
\begin{abstract}
We investigate Bayesian and frequentist approaches to resonance searches using a toy model based on an ATLAS search for the Higgs boson in the diphoton channel. We draw pseudo-data from the background only model and background plus signal model at multiple luminosities, from $10^{-3}\invfb$ to $10^7\invfb$. We chart the change in the Bayesian posterior of the background only model and the global \pvalue. We find that, as anticipated, the posterior converges to certainty about the model as luminosity increases. The \pvalue, on the other hand, randomly walks between $0$ and $1$ if the background only model is true, and otherwise converges to~$0$. After briefly commenting on the frequentist properties of the posterior, we make a direct comparison of the significances obtained in Bayesian and frequentist frameworks. We find that the well-known look-elsewhere effect reduces local significances by about $1\sigma$. We furthermore find that significances from our Bayesian framework are typically about $1\todash2\sigma$ smaller than the global significances, though the reduction depends on the prior, global significance and integrated luminosity. This suggests that even global significances could significantly overstate the evidence against the background only model. We checked that this effect --- the Bayes effect --- was robust with respect to \nprior choices of prior and investigated the Jeffreys-Lindley paradox for three of them.
\end{abstract}

\maketitle

\section{Introduction}

Resonances are narrow peaks in the invariant mass distributions of pairs of particles detected in, for example, a collider experiment. As a resonance originates from the on-shell production and subsequent decay of a massive particle, a new resonance indicates the discovery of a new particle, as was the case in 2012 when the Higgs boson was discovered\cite{Aad:2012tfa,Chatrchyan:2012xdj} at the Large Hadron Collider (LHC). New resonances typically lie above a background of events originating from well-understood physical processes. Unfortunately, upwards statistical fluctuations in that background can, by chance, imitate a resonance. Thus to discover a particle we must distinguish fluctuations from genuine resonances.

The ATLAS and CMS experiments at the LHC use null hypothesis significance testing to determine whether there is a new resonance in an invariant mass distribution. This methodology, reviewed in \refsec{sec:freq}, invokes hypothetical pseudo-experiments conducted with no resonance. If the fraction of pseudo-experiments that would result in data at least as extreme as that observed (the so-called \pvalue) lies below a threshold, we reject the possibility that there is no resonance, and possibly herald a discovery. In light of historical (see e.g., \refcite{ref1961-04282-001}) and recent (see e.g., \refcite{rss}) criticism, we recently advocated\cite{Fowlie:2016rmn,Fowlie:2017fya} an alternative methodology based on Bayes factors, reviewed in \refsec{sec:bayes}. In this methodology, we consider only the observed data, and directly compare the plausibility of that data in a model with a resonance to that in one without one. This allows us to calculate, amongst other things, the posterior probability of the background only model.\footnote{As discussed further in \refsec{sec:bayes}, we assume throughout that the background and signal models are a priori equally plausible.}

In \refcite{Fowlie:2016rmn} we considered the infamous $750\gev$ anomaly, when ATLAS and CMS saw hints of a resonance in the invariant mass distribution of photons at $750\gev$. This would have been a historic discovery and it generated considerable activity (see e.g., \refcite{Strumia:2016wys} for a review). Looking at ATLAS data\cite{TheATLAScollaboration:2015mdt,Aaboud:2016tru}, we found that whilst the \pvalue of about $0.02$ appeared suggestive, for our choices of priors the posterior of the background model was in fact about $0.2$. The anomaly ultimately disappeared once more data was collected\cite{ATLAS:2016eeo}. Similarly, in \refcite{Fowlie:2017fya}, we considered a hint for a resonance in the energy spectrum of electrons measured by the DAMPE satellite\cite{Ambrosi:2017wek}. The hint could be the first direct evidence for dark matter. We again found that for our choices of priors the posterior of the background, about $0.3$, was significantly greater than the \pvalue, about $0.01$.

With these two cases in mind, in this work we compare the behavior of the posterior of the background and the \pvalue in resonance searches. In \refsec{sec:toy} we present a toy example of a resonance search based on one in which ATLAS discovered the Higgs boson\cite{Aad:2013wqa}.
We review the frequentist and Bayesian methods for analyzing this toy problem in \refsec{sec:freq} and \refsec{sec:bayes}, respectively, and apply them to real and pseudo-data from it in \refsec{sec:results}, showing how results change as the dataset grows (see \refcite{PMID:26651986} for a similar analysis in a different context), and differences between the statistical approaches. Resonance searches are tests of point null hypotheses. Historically, such tests drew considerable attention and remain controversial (see 
e.g., \refcite{10.2307/2333251,doi:10.1080/01621459.1982.10477809,Bernardo1980,doi:10.1086/673730,berger1987,Trotta:2005ar,2013arXiv1303.5973R,Cousins:2013hry}) due to the Jeffreys-Lindley paradox\cite{Jeffreys:1939xee,10.2307/2333251}.

We find in \refsec{sec:bayes_effect} that the posterior of the background only model is typically orders of magnitude greater than the \pvalue.\footnote{\Refcite{johnson2013} reaches a similar conclusion in an extremely crude analysis of the Higgs discovery.}
This echoes criticisms that \pvalues overstate the evidence for effects in other fields (see e.g., \refcite{edwards1963bayesian,doi:10.1080/01621459.1987.10478397,rss}).
Despite arguments in e.g., \refcite{Lyons:2014pta,Senn2001TwoCF} that such comparisons are akin to comparing heights and weights, we feel that a comparison is justified as both the \pvalue (when interpreted in the manner of Fisher; see \refsec{sec:freq}) and posterior aim to answer the same question: how strong is the evidence against the null hypothesis? The fact that \pvalues markedly differ from results of Bayesian methods does not, however, imply that \pvalues are wrong.

We do not in this work address in detail further criticisms of either methodology, e.g., the dependence of Bayes factors on choices of prior (see e.g., \refcite{kass,Cousins:2008gf}), though discuss prior sensitivity of our findings in \refsec{sec:sensitivity} and the Jeffreys-Lindley paradox in \refsec{sec:jl}. We summarize our findings in \refsec{sec:conclusions}.

% \vfill  % Hack to remove big paragraph space

\section{Toy problem --- ATLAS diphoton search for a Higgs boson}\label{sec:toy}

\begin{figure}[t]
    \centering
    \includegraphics[width=\figwidth]{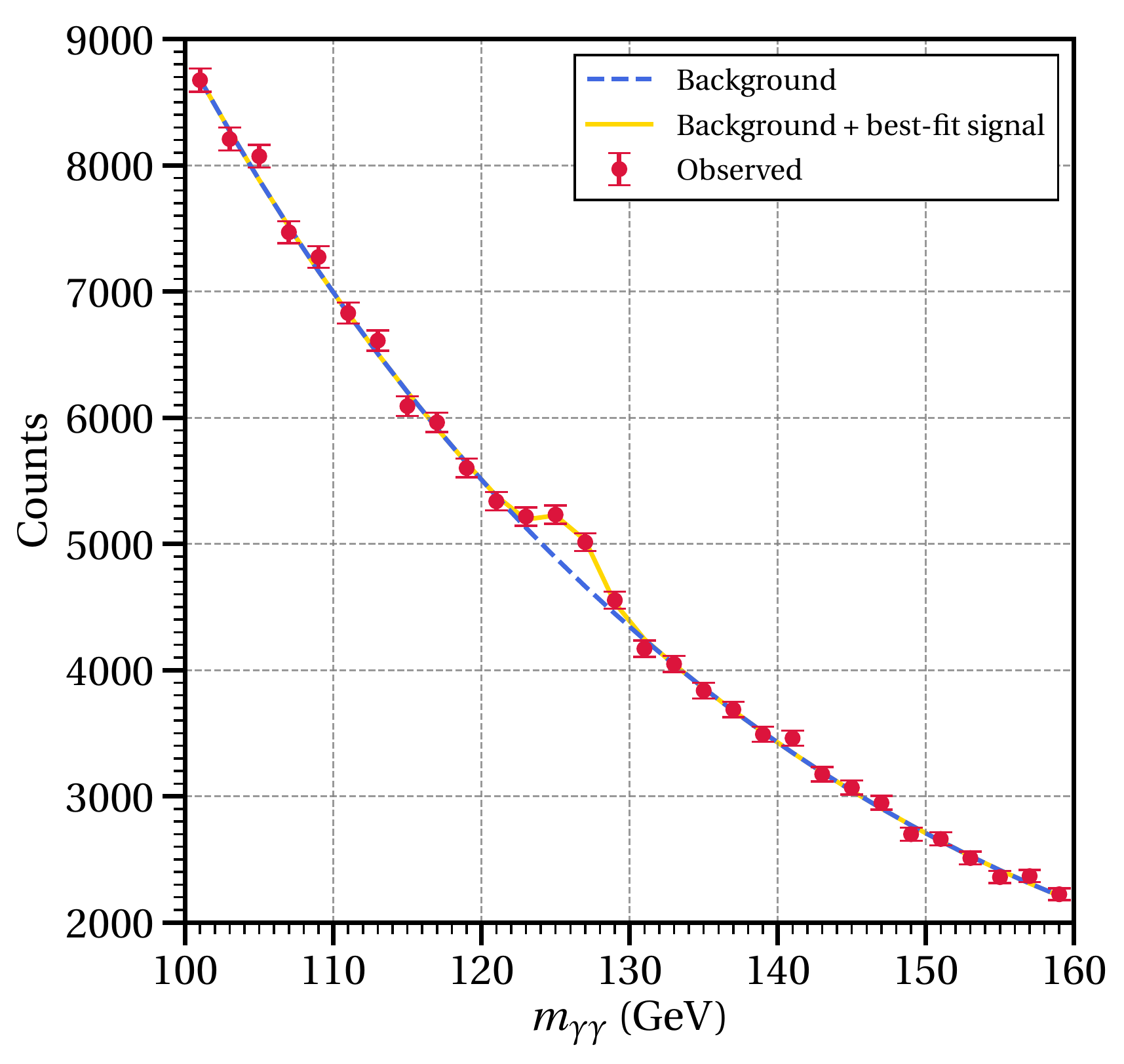}
    \caption{Diphoton spectrum observed by ATLAS\cite{Aad:2013wqa} (red points), predicted by the background only model (dashed blue) and by the best-fitting signal model plus background (solid yellow).}
    \label{fig:spectrum}
\end{figure}

We consider a historic ATLAS\cite{Aad:2013wqa} search for a Higgs boson in the diphoton channel at $\sqrt{s}=7\tev$ and $\sqrt{s}=8\tev$ with $25\invfb$ of integrated luminosity; this search contributed to the discovery of the Higgs boson in 2012. The inclusive observed spectrum is shown in \reffig{fig:spectrum}. The spectrum (red points) exhibits a suggestive resonance-like feature at $125\gev$. We consider two models for the data.

First, a background only model, $M_0$, representing the background from known Standard Model (SM) processes other than a Higgs boson. ATLAS model the background by a Bernstein polynomial. This introduces several unknown coefficients and an unknown normalization that governs the overall expected background yield. So that we can perform millions of fits, we fix the coefficients and normalization to their best-fit values and thus omit all systematic uncertainties in the background, including parametric uncertainties in the Bernstein coefficients and uncertainties associated with that particular choice of functional form.

Second, a background plus signal model, $M_1$. This is the background spectrum plus a narrow resonance, i.e., one with a width, $\Gamma$, that is substantially narrower than the experimental resolution of about $1.5\gev$. We thus approximate the width as zero, $\Gamma \approx 0$. We know the expected amplitude of a resonance from an SM-like Higgs boson. We choose to model a Higgs-like signal with only two a priori unknown parameters: the mass of the Higgs, $m_h$, and the amplitude of the signal relative to the SM, $\mu\ge0$.

The observed data is a set of counts, $o_i$ with $i=1,2,\dotsc, 30$, in each of $30$ bins in diphoton invariant mass spanning $m_{\gamma\gamma} = 100 \todash 160\gev$. We denote the expected number of background events per bin by $b_i$. In the signal model, the expected number of events is the sum of the expected number of background events and signal events, $s_i$,
\begin{equation}
\lambda_i = b_i + \mu s_i(m_h).
\end{equation}
The expected signal is scaled by the signal strength, $\mu\ge0$, and itself depends on the Higgs mass by
\begin{equation}
s_i(m_h) = \sigma \times \epsilon \times \lumi \times \int_{m_i - \tfrac12 \Delta m}^{m_i + \tfrac12 \Delta m} \Gaussian\left(m_{\gamma\gamma}; m_h\right) \dif m_{\gamma\gamma},
\end{equation}
where $m_i$ is the bin center, $\Delta m \equiv 2\gev$ is the bin width, $\lumi = 25\invfb$ is the integrated luminosity, $\sigma$ is the total production cross section, $\epsilon$ is the selection efficiency, and $\phi$ is the
shape of the signal. We model the shape by a Gaussian centered at the Higgs mass, $m_h$, with a width of $1.5\gev$, governed by the experimental resolution. We pick a Gaussian for simplicity; ATLAS in fact use a modified Crystal Ball function. We found $\sigma \times \epsilon \simeq 22.3\fb$ heuristically by reproducing the ATLAS result at $m_h = 125\gev$. The total background cross section, on the other hand, was about $5700\fb$.

Thus the likelihood of the observed events is the product of $30$ Poisson likelihoods, one for each bin,
\begin{equation}\label{eq:like}
\Prob{o}{\lambda} = \prod_{i=1}^{30} \frac{e^{-\lambda_i} \lambda_i^{o_i}}{o_i!},
\end{equation}
where $\lambda_i$ and $o_i$, are the expected and observed numbers of events in bin $i$, respectively. We stress that this is a toy treatment of the problem, intended to capture only the major details, that omits e.g., systematic uncertainties or a detailed treatment of the selection efficiency and cross section as functions of the Higgs mass.

\section{Frequentist}\label{sec:freq}

In the frequentist framework, we construct a log likelihood ratio test-statistic,
\begin{equation}\label{eq:test_statistic}
q \equiv 2\ln\left(\frac{\Prob{\data}{\hat\mu, \hat{m}_h, M_1}}{\Prob{\data}{M_0}}\right),
\end{equation}
where $\data$ represents our experimental data and the hats indicate the best-fitting value of a parameter. The best-fitting signal strength cannot be negative, $\hat\mu\ge0$. Using the likelihood in \refeq{eq:like}, we have that
\begin{align}
\Prob{\data}{\mu, m_h, M_1} &= \Prob{o}{\lambda = b + \mu s(m_h)},\\
\Prob{\data}{M_0} &= \Prob{o}{\lambda = b}\label{eq:like_background}.
\end{align}
The frequentist quantities of interest are functions only of $q$, which itself implicitly depends on $\hat\mu$ and $\hat{m}_h$. To ensure that no signals are missed, we find the best-fit mass and signal strength inside every bin with an excess by Brent's method\cite{scipy,brent}, using $(o - b) / s$ for that bin as a starting guess for the signal strength. Lastly, we select the best-fit mass and signal strength found from all the bins.

We use the test-statistic to calculate \pvalues with respect to a null hypothesis, which in this work is always the background only model.
The global \pvalue is the probability of obtaining a test-statistic at least as extreme as
that observed for any value of the Higgs mass, assuming that the background only model is true,
\begin{equation}
\pvalue = \Prob{q \ge q_\text{Observed}}{M_0}.
\end{equation}
A local \pvalue, on the other hand, presupposes that we were testing only a specific value of the Higgs mass. The \pvalue is distributed as uniformly as possible under the background only model\cite{cdf}. Thus, if we reject the background model only when the \pvalue lies below a threshold specified in advance,
\begin{equation}
\text{If $\pvalue < \alpha$, reject $M_0$},
\end{equation}
the threshold $\alpha$ is the type-1 error rate, i.e., the probability of rejecting the background model when it is true.
The \pvalue itself though has no frequentist or Bayesian meaning: it is not an error rate or the probability of the background only model\cite{doi:10.1080/00031305.1975.10479106}. In high-energy physics, \pvalues are informally interpreted in the manner of Fisher\cite{fisher} as a continuous measure of evidence against the background only model, though discoveries traditionally require a \pvalue below about $10^{-7}$, corresponding to $5\sigma$\cite{Lyons:2013yja}.\footnote{See \refcite{doi:10.1198/0003130031856} for a discussion of this hybrid interpretation.}

To calculate \pvalues, we make use of two formulas that are appropriate because our test-statistic is a log likelihood ratio between nested hypotheses (the signal plus background model contains the background model at $\mu=0$). For local \pvalues, we use an asymptotic formula\cite{Cowan:2010js} based on Wilks' theorem\cite{Wilks:1938dza,chernoff1954}
\begin{equation}\label{eq:local_p}
\text{Local \pvalue} = 1 - \Phi\left(\sqrt{q}\right),
\end{equation}
where $\Phi$ is the cumulative distribution function of a standard normal distribution. This presupposes that we were testing only the best-fitting Higgs mass, $\hat{m}_h$.
For global \pvalues, we try a bootstrap technique: as discussed in \refsec{sec:results} we ultimately perform \npseudo pseudo-experiments at several luminosities, repeating the minimization in \refeq{eq:test_statistic} each time.
We count the fraction of pseudo-experiments that result in a test-statistic greater than or equal to that observed.\footnote{%
In other words, we recycle our pseudo-experiments. First, we use them to simulate the distribution of the test-statistic under the null hypothesis. Second, we use them in our comparisons of Bayesian and frequentist techniques.
}
If the resulting global \pvalue lies below $0.001$, we make use of the Gross-Vitells technique\cite{Gross:2010qma}, which should be more precise than bootstrap in that regime,
\begin{equation}
\pvalue \approx \tfrac12 \Prior{\chi^2_1 > q} + N e^{-q/2},
\end{equation}
where $\chi^2_1$ is a random variate from a chi-squared with one degree of freedom and $N$ is an a priori unknown parameter.\footnote{%
Under the asymptotic conditions required by the Gross-Vitells method, $N$ doesn't depend on the luminosity.} We determined $N$ by counting the number of upcrossings of the test-statistic between $100\gev$ and $160\gev$ in \npseudo Monte-Carlo (MC) simulations.
Thus this formula accounts for a one-dimensional look-elsewhere effect for the Higgs mass in the range $100\gev$ to $160\gev$.\footnote{%
Note that ATLAS\cite{Aad:2012tfa} corrected for look-elsewhere effects in the ranges $110\todash600\gev$ and $110\todash150\gev$,
whilst CMS\cite{Chatrchyan:2012xdj} corrected for them in the ranges $115\todash130\gev$ and $110\todash145\gev$. The narrower ranges reflected indirect constraints and exclusion limits from previous experiments and mitigated edge effects at the boundaries of the data. We checked that edge effects were absent by checking that the distribution of the log-likelihood ratio matched the expected asymptotic result\cite{Cowan:2010js} at the boundaries. We furthermore checked that narrowing the interval to $110\todash150\gev$ did not qualitatively affect our results.
}
This was unaccounted for by \refeq{eq:local_p} because the likelihood's dependence on the Higgs mass vanishes on the $\mu = 0$ boundary, breaking an assumption of Wilks' theorem.

Finally, we convert \pvalues to \zvalues, as is conventional, using a one-tailed Gaussian rule,
\begin{equation}\label{eq:z_value}
Z = \Phi^{-1}(1 - p).
\end{equation}

\section{Bayesian model comparison}\label{sec:bayes}

% This is nothing other than the direct application of probability theory.
We consider the probability of the data in each model --- the so-called Bayesian evidence. For the signal model, we must marginalize the a priori unknown Higgs mass and signal strength,
\begin{equation}\label{eq:ev}
\Prob{\data}{M_1} = \int \Prob{\data}{m_h, \mu} \prob{m_h, \mu}{M_1} \dif m_h \dif\mu,
\end{equation}
where the second factor in the integrand is our choice of prior density for the Higgs mass and signal strength. We perform the integration by adaptive quadrature\cite{quadpack,scipy}, treating the bin centers of the five bins with the most significant excesses as special points.
For the background model, it is trivial, as there are no free parameters, so the evidence is given directly by \refeq{eq:like_background}. 

Clearly, the evidence for the signal model depends upon our choice of prior.
For the mass, we focused on the narrow range searched by ATLAS of $100\gev$ to $160\gev$ --- although choices that extend this range could reflect prior belief, they would simply dilute the evidence for the signal model.
Since it did not span multiple decades, we picked a flat prior. For the signal strength, we imagined a scenario in which we expected it to be somewhat close to the SM prediction of $\mu=1$, though could deviate by a couple of orders of magnitude. We supposed that we were ignorant of the magnitude of this deviation and thus picked a prior that was flat in logarithmic space between $10^{-2}$ and $10^2$. This prior does not favor any particular order of magnitude, $\prior{\ln \mu}=\text{const.}$; however, the prior predictive for the best-fit signal strength could favor particular magnitudes depending on the likelihood.
We investigate several other choices of prior in \refsec{sec:sensitivity}.

With the evidences, we calculate the Bayes factor (see e.g., \refcite{kass} for a review),
\begin{equation}\label{eq:bayes_factor}
B_{10} = \frac{\Prob{\data}{M_1}}{\Prob{\data}{M_0}}.
\end{equation}
This is the impact of the data on the relative plausibility of models $M_1$ and $M_0$. As the luminosity grows, we expect the Bayes factor to increasingly favor the true model\cite{doi:10.1080/00031305.2017.1397548, 2016arXiv160700292C}. To facilitate a comparison with \pvalues, we find the posterior of the background model,
\begin{equation}\label{eq:posterior}
\begin{split}
\text{Posterior of background} &\equiv \Prob{M_0}{\data}\\
&= \frac{\Prior{M_0}}{\Prior{M_0} + B_{10} \Prior{M_1}}.
\end{split}
\end{equation}
We can coherently incorporate relevant background information in our choices of priors for the models, $P(M_0)$ and $P(M_1)$.
From hereon we assume that the models are a priori equally plausible, $\Prior{M_0} = \Prior{M_1}$, though if we had a reason for considering new physics relatively implausible, we could choose $\Prior{M_0} \gg \Prior{M_1}$.
In this work we compute posteriors to enable a comparison with \pvalues; however, we usually advocate reporting only Bayes factors, since they are independent of the priors for the models themselves.

We, furthermore, convert the posterior into a \zvalue\ --- which we call the Bayesian significance --- using the rule in \refeq{eq:z_value}. We do not advocate this in general and do so only to communicate our results to particle physicists so used to thinking of anomalies in terms of the number of sigmas.

There exists a bound on the rate at which we would be misled by the posterior about the true model\cite{kerridge1963}. With two models under consideration that are equally plausible a priori, if model $M_0$ is correct, the chance that we find a posterior for $M_0$ that is less than a threshold $t$ is bounded by
\begin{equation}\label{eq:kerridge}
\Prob{\subbox{$\Prob{M_0}{\data} \le t$}}{M_0} \le \frac{t}{1 - t}.
\end{equation}
The general bound depends on the number of models considered and the prior odds between the models.

For our toy problem, there exists a lower bound on the posterior of the background obtainable with any priors for the mass and signal strength (see \refcite{delampady1990} and references therein for further discussion of lower bounds on the posterior). The posterior is minimized when the priors select the best-fitting mass and strength, such that
\begin{equation}
\Prob{M_0}{\data} = \frac{1}{1 + e^{q / 2}},
\end{equation}
where $q$ is the test-statistic defined in \refeq{eq:test_statistic}.

There are two major causes of discrepancies between posteriors and \pvalues. First, the fact that the \pvalue considers the probability that the test-statistic exceeds a threshold, i.e, the probability of
\begin{equation}\label{eq:E}
E \equiv q \ge q_\text{Observed},
\end{equation}
rather than the probability of obtaining the observed data or observed test-statistic. To aid our comparisons between Bayesian and frequentist procedures, we thus compute the posterior of the background assuming this coarse-grained information, i.e., we calculate
\begin{equation}
\text{Coarse-grained posterior of background} \equiv \Prob{M_0}{E}.
\end{equation}
In the Bayesian context, conditioning upon $E$ is strange as we should condition upon all that we know, i.e., the observed data itself. There is a gulf between knowing the data and knowing only that the data was in the set $E$ (see \refcite{berger1987} for further discussion). This is useful though as if the models are equally plausible a priori,
\begin{equation}\label{eq:posterior_extreme}
\Prob{M_0}{E} = \frac{\pvalue}{\pvalue + \Prob{E}{M_1}},
\end{equation}
where $\Prob{E}{M_1}$ may be calculated using asymptotic formulae developed for frequentist applications. When the \pvalue is small, \refeq{eq:posterior_extreme} can be written as,
\begin{equation}
\Prob{M_0}{E} \approx \frac{\pvalue}{1 - \Prob{\notE}{M_1}}.
\end{equation}
In this form it resembles \CLs\cite{Junk:1999kv,Read:2002hq} and Birnbaum's measure of evidence in a binary experiment\cite{zbMATH03192484,doi:10.1080/01621459.1962.10480660}.
Under the signal model, $\Prob{E}{M_1}$ is distributed as uniformly as possible, such that its median is about one half. Thus for small \pvalues the median coarse-grained posterior approximately equals twice the \pvalue.

The second major cause of discrepancies between posteriors and \pvalues is the so-called Occam effect\cite{Mackay}. The evidence in \refeq{eq:ev} may be thought of as the likelihood averaged upon the prior. If the prior typically poorly predicts the observed data by e.g., predicting signal strengths much greater than favored by data, the evidence would be automatically penalized by averaging, even if the prior permits a signal strength that successfully predicts the observed data. This means that the posterior for the signal model would typically be reduced by broadening the prior range for e.g., the signal strength\cite{10.1093/biomet/44.3-4.533}.

% This effect may be quantified by the Occam factor, which is defined as the Bayes factor divided by the likelihood ratio and in this context may be written,
% \begin{equation}
% \mathcal{O} = B_{10} e^{-q/2}
% \end{equation}
% where the test-statistic $q$ was defined in \refeq{eq:test_statistic}.

\begin{figure*}[t]
    \centering
    \includegraphics[width=\widefigwidth]{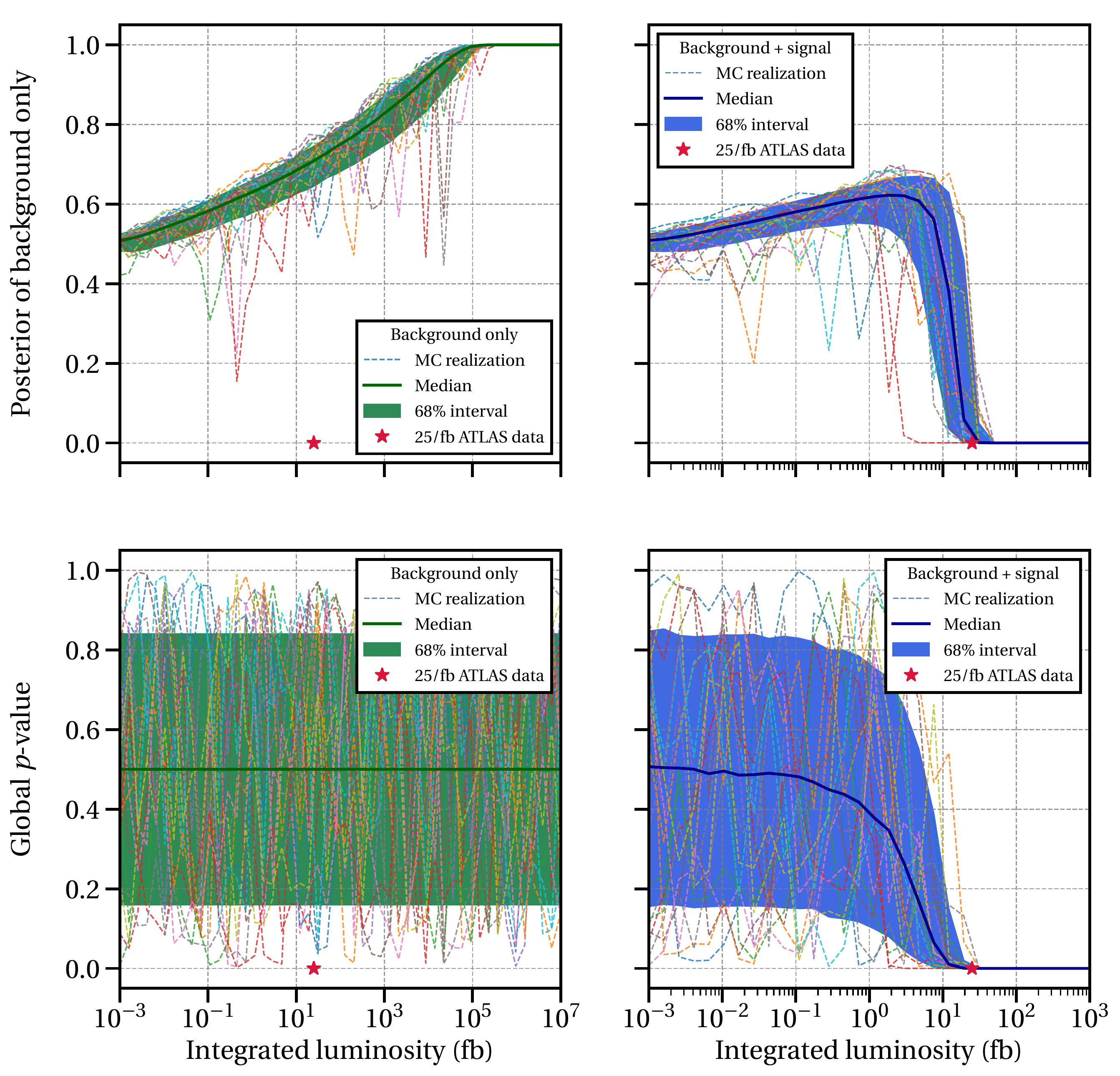}
    \caption{Changes in the median posterior of the background (top) and global \pvalue (bottom) as the integrated luminosity grows assuming the background only model (left) and the signal model with $\mu=1$ and $m_h=125\gev$ (right). The band shows an equal-tailed $68\%$ interval. A random subset of MC realizations are shown with dashed lines. The \pvalue and posterior corresponding to data observed by ATLAS (red stars) lie extremely close to zero in all cases.}
    \label{fig:cumulative}
\end{figure*}

\subsection{The Jeffreys-Lindley paradox}\label{sec:intro_jl}

The Jeffreys-Lindley paradox\cite{Jeffreys:1939xee,10.2307/2333251} demonstrates that a posterior and a \pvalue could result in opposite conclusions even in the asymptotic limit. The paradox results from the fact that the posterior corresponding to a particular \pvalue may depend on the sample size (i.e., integrated luminosity in the context of collider physics) such that the disagreement between the posterior and \pvalue could diverge as the sample size increases. The paradox prompted questions about the foundations of hypothesis testing; for commentary and attempted resolutions, see e.g., \refcite{10.2307/2333251,doi:10.1080/01621459.1982.10477809,Bernardo1980,doi:10.1086/673730,berger1987,Trotta:2005ar,2013arXiv1303.5973R,Cousins:2013hry}.

In the example of the paradox in \refcite{Cousins:2013hry} we imagine $n$ Gaussian measurements of an unknown parameter, $\mu$, with a sample mean $\bar x \sim \mathcal{N}(\mu, \sigma^2 /n)$. We consider two hypotheses: the null $H_0$ in which $\mu = 0$ and an alternative $H_+$ in which $\mu \ge 0$. The sample mean is a sufficient statistic such that evidence for the null is simply a Gaussian evaluated at $\mu = 0$,
\begin{equation}
\prob{\bar x}{H_0} = \prob{\bar x}{\mu = 0} = \tfrac{1}{\sqrt{2\pi}} \tfrac{\sqrt{n}}{\sigma} e^{-\tfrac{1}{2} \chi^2},
\end{equation}
where $\chi^2 = n \bar x^2 / \sigma^2$ is the chi-squared. For the evidence for the alternative model, we must average the likelihood upon a prior for the unknown parameter, $\prior{\mu}$,
\begin{equation}
\begin{split}
\prob{\bar x}{H_+} &= \int \prob{\bar x}{\mu} \prior{\mu} \dif \mu\\
&= \int \tfrac{1}{\sqrt{2\pi}} \tfrac{\sqrt{n}}{\sigma} e^{- \tfrac{n}{2} (\bar x - \mu)^2 / \sigma^2} \prior{\mu} \dif \mu.
\end{split}
\end{equation}
The Bayes factor is given by their ratio,
\begin{equation}\label{eq:bf_jl}
B_{+0} =%
\frac{
\int \tfrac{1}{\sqrt{2\pi}} \tfrac{\sqrt{n}}{\sigma} e^{- \tfrac{n}{2} (\bar x - \mu)^2 / \sigma^2} \prior{\mu} \dif \mu%
}%
{%
\tfrac{1}{\sqrt{2\pi}} \tfrac{\sqrt{n}}{\sigma} e^{-\tfrac{1}{2}\chi^2}%
}%
.
\end{equation}
As $n$ grows the exponential term in the integrand becomes sharply peaked at $\mu = \bar x$, such that we may make a Laplace approximation resulting in
\begin{equation}\label{eq:bf_jl_approx}
B_{+0} \approx e^{\tfrac12\chi^2} \prior{\mu = \bar x} \frac{\sqrt{2\pi}\sigma}{\sqrt{n}}.
\end{equation}
The \pvalue, on the other hand, depends only on the chi-squared by $\pvalue = 1 - \Phi(\chi)$ for a one-sided test.

To create the paradox, we consider fixed chi-squared but increasing sample size. If $\prior{\mu = \bar x}$ is independent of $n$, from \refeq{eq:bf_jl_approx} we see that for fixed chi-squared the Bayes factor tends to zero as the sample size increases. Thus, despite an arbitrarily small \pvalue, the Bayes factor could overwhelmingly favor the null. To maintain a fixed chi-squared, however, the sample mean must scale as $\bar x = \chi \sigma / \sqrt{n}$. This reflects the fact as we accumulate data we are sensitive to smaller effects.

\section{Results}\label{sec:results}

\subsection{Trajectories as we collect luminosity}\label{sec:trajectories}

We begin in \reffig{fig:cumulative} by investigating the change in the \pvalue and posterior of the background as the integrated luminosity grows.
First, we sampled data from the background only model, performing \ncumulative pseudo-experiments at \nlumi luminosities between $10^{-3}\invfb$ and $10^7\invfb$. As we increased the luminosity, we added events to existing datasets, such that they were correlated. At each luminosity,
we found the median and an equal-tailed $68\%$ interval for the posterior of the background model and the \pvalue. We found that the posterior increases monotonically from $0.5$ to $1$ as the luminosity increases, with a narrow $68\%$ interval (top left panel), i.e., as the luminosity grows, the true model is increasingly favored. The rate of convergence, however, is slow; it takes about $10^5\invfb$ before any strong preference for the background emerges. The \pvalue, on the other hand, makes a random walk, with a median of $0.5$ regardless of the luminosity (bottom left panel).

\begin{figure}[t]
    \centering
    \includegraphics[width=\figwidth]{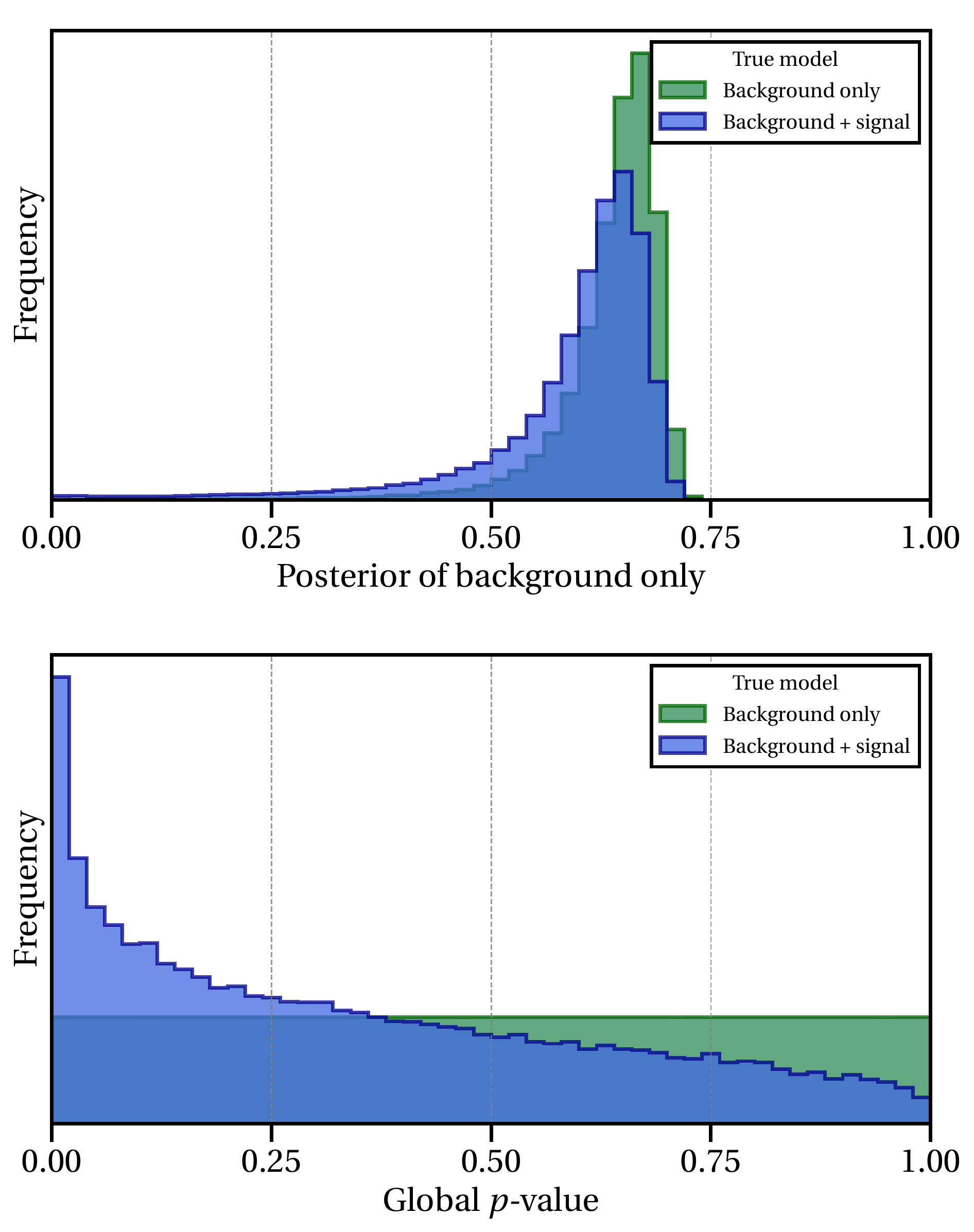}
    \caption{Distributions of posterior of the background (top) and global \pvalue (bottom) with $2.5\invfb$ assuming the background only model (green) and the background + signal model with $\mu=1$ and $m_h=125\gev$ (blue), which predicts about $56$ signal events.}
    \label{fig:pseudo_2.5}
\end{figure}

\begin{figure}[t]
    \centering
    \includegraphics[width=\figwidth]{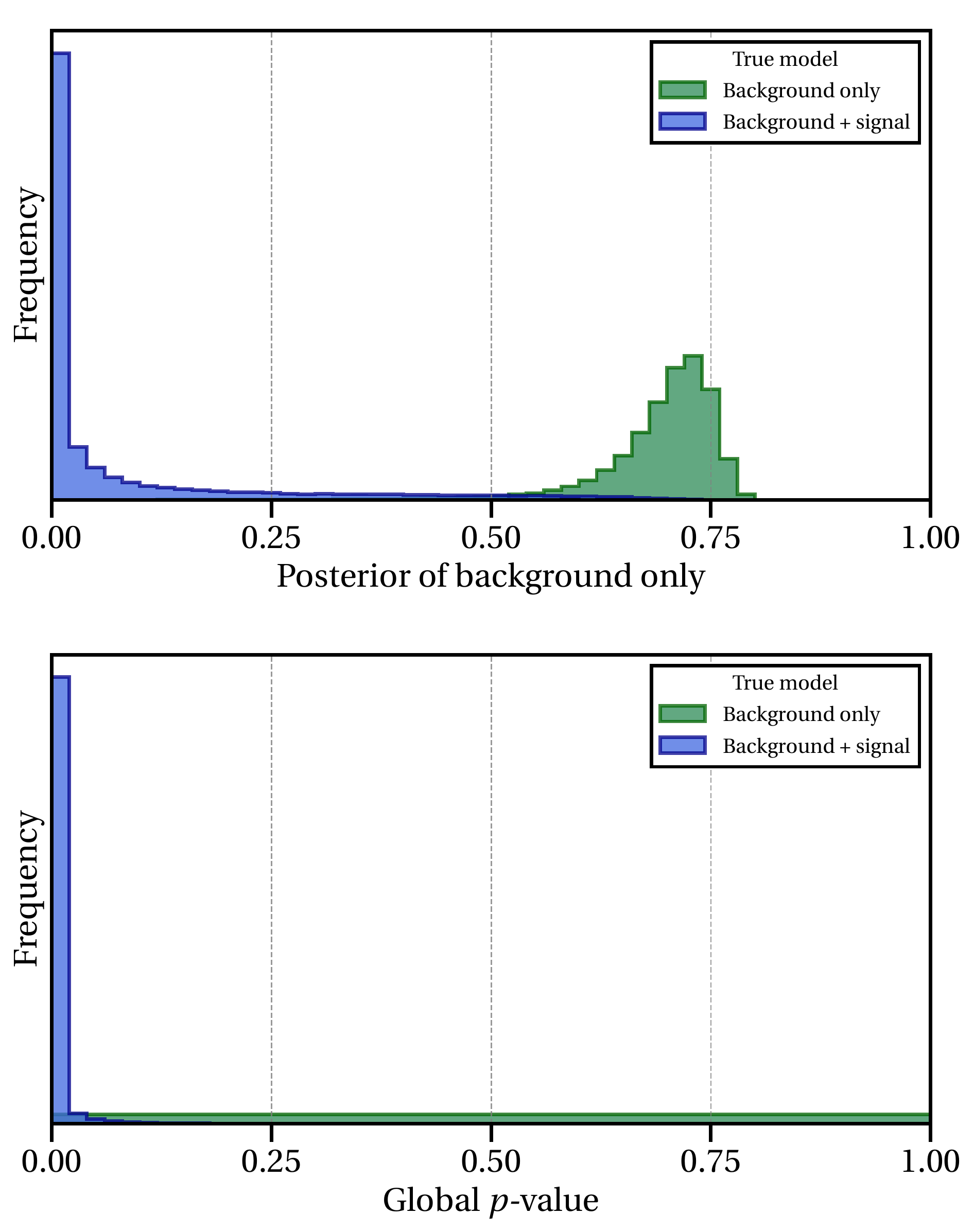}
    \caption{As in \reffig{fig:pseudo_2.5} but with $25\invfb$. The background + signal model predicts about $560$ signal events.}
    \label{fig:pseudo_25}
\end{figure}

Second, we sampled data from the signal model assuming a signal strength $\mu=1$ and a mass $m_h=125\gev$, performing \ncumulative pseudo-experiments at the same \nlumi luminosities. Surprisingly, the posterior first increases from $0.5$, before rapidly decreasing to $0$ once the luminosity exceeds about $10\invfb$ (top right panel). The initial increase in the posterior is, however, mild; the median reaches about $0.6$, which shouldn't be considered worth more than a bare mention\cite{Jeffreys:1939xee}.
It occurs because below about $10\invfb$ Poisson fluctuations in the background of order $\sqrt{b}$ provide an economical explanation of a signal-like feature as $s \approxless\sqrt{b}$. The signal model isn't strongly rewarded as it cannot provide a much better fit to the data; however, it is punished by the Occam effect as it wastes prior probability at $\mu \gg 1$, which would typically result in signals that were not observed. This results in a mild preference for the background model. Above about $10\invfb$, fluctuations in the background cannot explain a signal as $s \gg \sqrt{b}$, and thus the posterior of the background decreases to $0$, as expected. In other words, at large luminosities, the reward from a better fit overcomes the Occam effect.

The \pvalue, on the other hand, decreases monotonically to $0$ (lower right panel) though within a substantial $68\%$ interval. As the signal and background models are similar at small luminosities, the interval for the \pvalue at small luminosities is similar to that when the background model is true. We show a random subset of MC realizations (dashed lines), illustrating possible trajectories of the \pvalue and posterior.

\subsection{Sampling properties of \pvalues and posteriors}

We continue by performing \npseudo pseudo-experiments under the background only model and signal model assuming a signal strength $\mu=1$ and a mass $m_h=125\gev$, at integrated luminosities of $0.25\invfb$, $2.5\invfb$, $12.5\invfb$, $25\invfb$, $50\invfb$ and $250\invfb$, though this time the datasets are completely independent. In \reffig{fig:pseudo_2.5} we show the resulting distributions of the posterior and \pvalue at $2.5\invfb$. The posterior peaks at about $0.6$ under the signal and background models, though under the latter mildly greater posteriors are favored.
Thus, even when the signal model is true, the posterior typically mildly favors the background model.
The explanation is identical to that in \refsec{sec:trajectories}: below about $10\invfb$, a fluctuation in the background can mimic the $\mu=1$ signal whereas the signal model wastes prior probability at $\mu \gg 1$.
The \pvalue on the other hand is flat under the background distribution, but peaks near $0$ under the signal model, with a substantial tail extending to $1$. At $25\invfb$, shown in \reffig{fig:pseudo_25}, the posterior under the background model peaks at about $0.75$, with greater preference rare, and under the signal model, it peaks near $0$, with a moderate tail. The \pvalue of course remains flat under the background distribution, but now sharply peaks near $0$ under the signal model.

\subsection{Type-1 errors and power}

\begin{figure}[t]
    \centering
    \includegraphics[width=\figwidth]{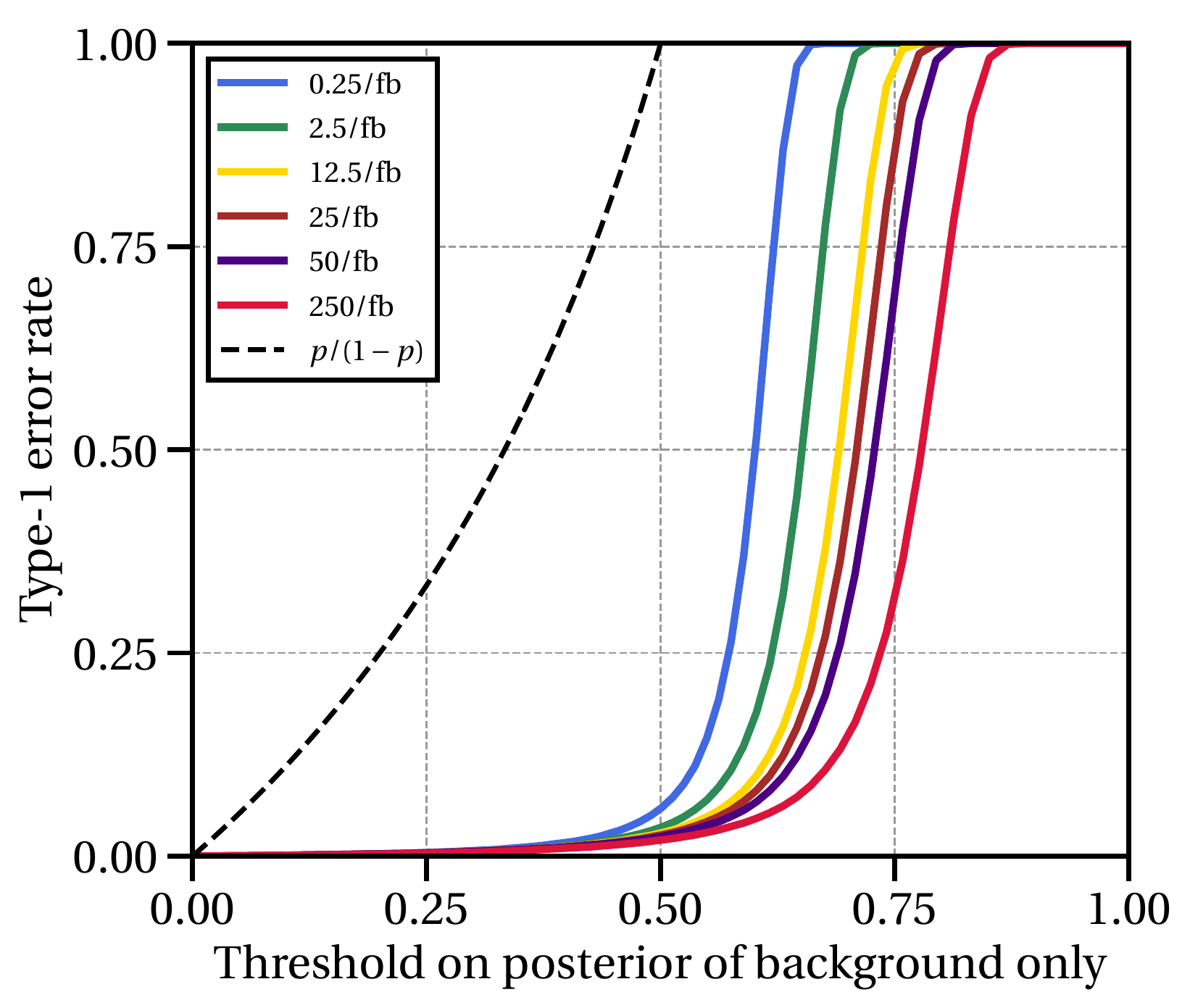}
    \caption{Type-1 error associated with a threshold on the posterior of the background only model for different integrated luminosities (solid lines) and the upper bound in \refeq{eq:kerridge} (dashed line).}
    \label{fig:type_1}
\end{figure}

In \reffig{fig:type_1} we consider the type-1 error associated with the posterior, i.e., if we placed a threshold on the posterior, what is the type-1 error rate? We see that this depends on the luminosity, though for all luminosities below posteriors of about $0.5$, the type-1 error rate is substantially less than the posterior and the bound in \refeq{eq:kerridge}. Thus, although in the Bayesian framework we relinquish direct control over the type-1 error rate, we find no evidence that a threshold on the posterior would lead to significant type-1 error rates in this setting.\footnote{We do not, however, necessarily recommend making a dichotomous decision, based on a threshold on the posterior or anything else.}

\begin{figure}[t]
    \centering
    \includegraphics[width=\figwidth]{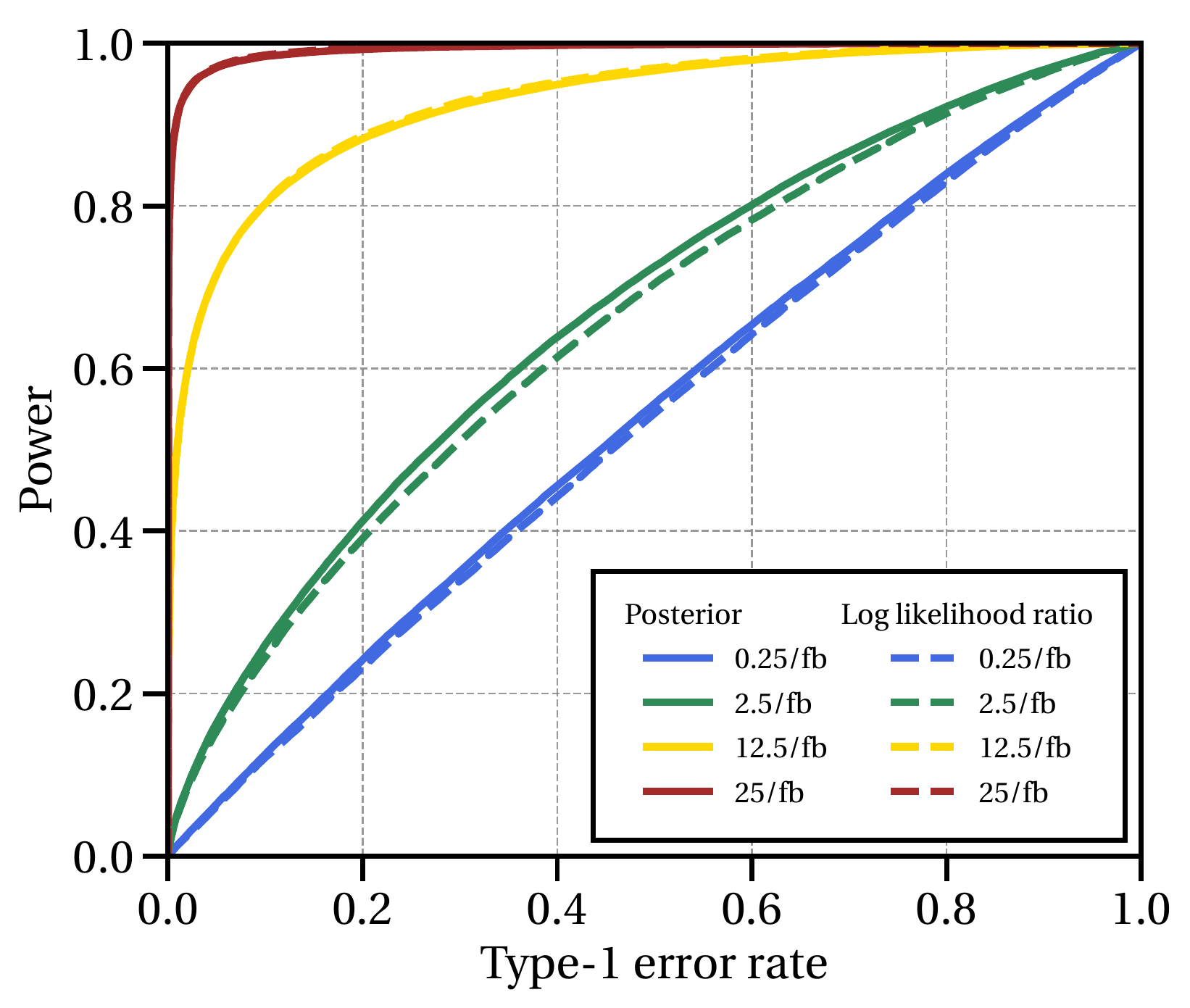}
    \caption{ROC curves at several luminosities (shown by different colors) using the posterior (solid lines) and log likelihood ratio (dashed lines) as a test-statistic.}
    \label{fig:ROC}
\end{figure}

In a similar manner, we investigated power --- the probability of rejecting the background model when we generated data from the signal model with $\mu=1$ and $m_h=125\gev$. We show in \reffig{fig:ROC} that the ROC curves --- the power versus the type-1 error rate --- were extremely similar for the log likelihood ratio in \refeq{eq:test_statistic} and the posterior in \refeq{eq:posterior}. They would thus result in similar type-2 error rates for a fixed type-1 rate when used as test-statistics in a hypothesis test. This is not surprising --- it is consistent with the fact that we observed an approximately monotonic relationship between the posterior and log likelihood ratio in this setting and the fact that error rates are invariant under strictly monotonic transformations of the test-statistic.

% The type-2 error rates that result from placing a threshold on the posterior are significantly greater than those from placing the same threshold on the \pvalue. That is, the posterior at first glance appears to have limited power to reject the background model when it is false. It shouldn't surprise us, however, that the power was limited since the type-1 error rate was substantially smaller than the threshold.

\subsection{The Bayes effect}\label{sec:bayes_effect}

\begin{figure}[t]
    \centering
    \includegraphics[width=\figwidth]{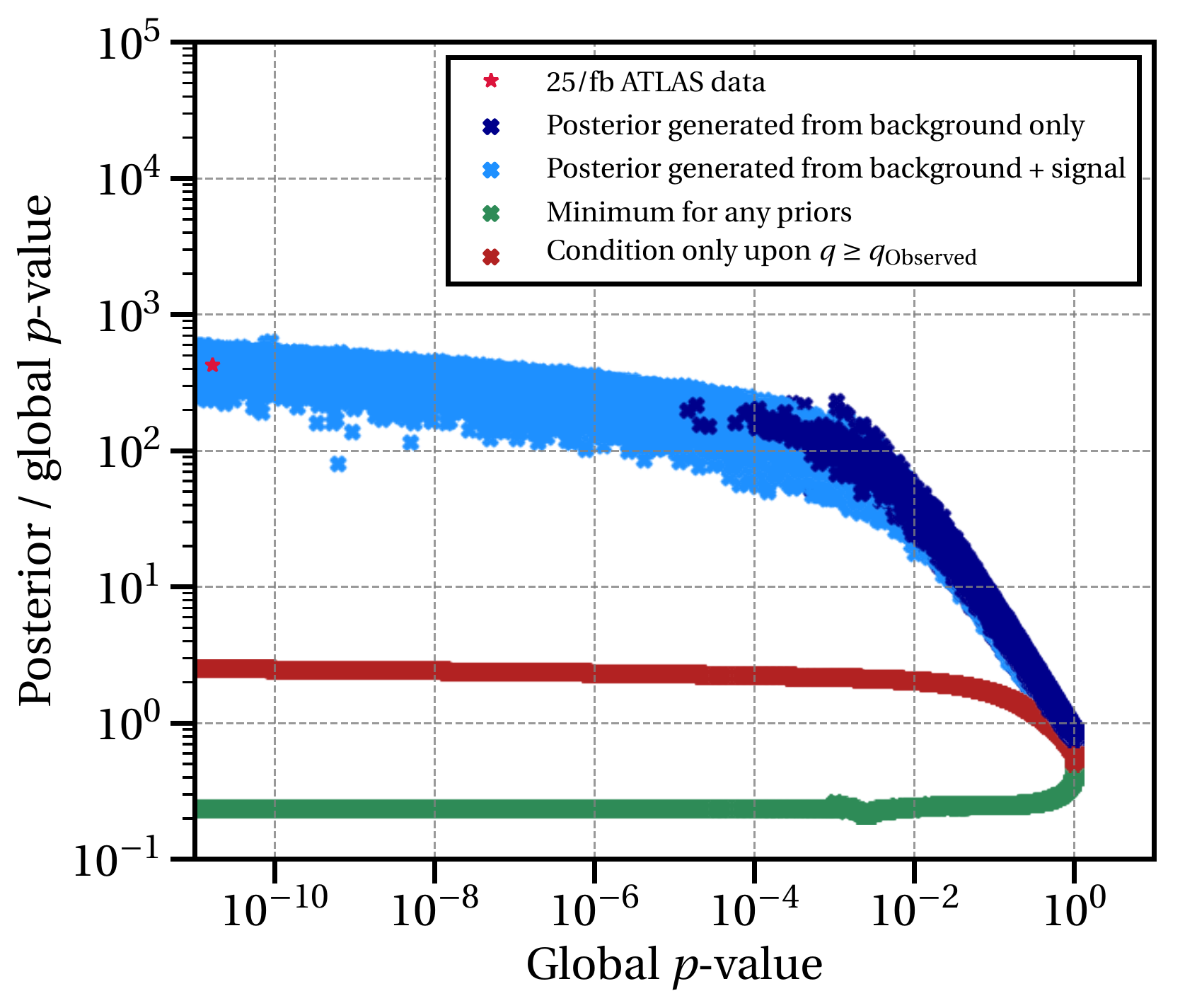}
    \caption{The ratio of the posterior of the background only model to \pvalue scattered against the \pvalue (blue points). We furthermore show the results corresponding to data observed by ATLAS (red star), the minimum possible posterior (green points) and that found by conditioning upon only the fact that $q \ge q_\text{Observed}$ (red points). Results from data drawn from the background only model are shown in darker colors.}
    \label{fig:difference}
\end{figure}

In \reffig{fig:difference}, we contrast the results from Bayesian and frequentist approaches. The ratios between the posterior of the background only model and \pvalue are scattered against the \pvalues from \npseudo pseudo-experiments at $0.25\invfb$, $2.5\invfb$, $12.5\invfb$, $25\invfb$, $50\invfb$ and $250\invfb$ (blue points). We do not find any significant differences in the relationship between \pvalue and posterior at different luminosities and thus scatter them together.\footnote{One might, however, have anticipated a Jeffreys-Lindley-style paradox\cite{10.2307/2333251,Jeffreys:1939xee}. The effect does not occur with a logarithmic prior; see \refsec{sec:jl} for further discussion.} The global \pvalues are significantly smaller than the associated posterior of the background by about $100\todash1000$ for \pvalues smaller than about $0.001$. Whilst both are extremely small, the global \pvalue from the data observed by ATLAS is about $400$ times smaller than the posterior (red star). We furthermore show that the minimum posterior obtainable with any priors for the Higgs mass and signal strength lies slightly below the global \pvalue (green points). We return to the issue of priors in \refsec{sec:sensitivity}.

Lastly, we show the effect of conditioning only upon the fact that the test-statistic was greater than the observed test-statistic, \refeq{eq:E}. This puts the exact experimental information used in the calculation of the \pvalue into the Bayesian one. We see that in this case the posteriors and \pvalues (red points) are similar, suggesting that this --- and not the Occam effect, prior information or differences in interpretation of probability --- may be the fundamental origin of the differences between the posteriors and \pvalues.

\begin{figure}[t]
    \centering
    \includegraphics[width=\figwidth]{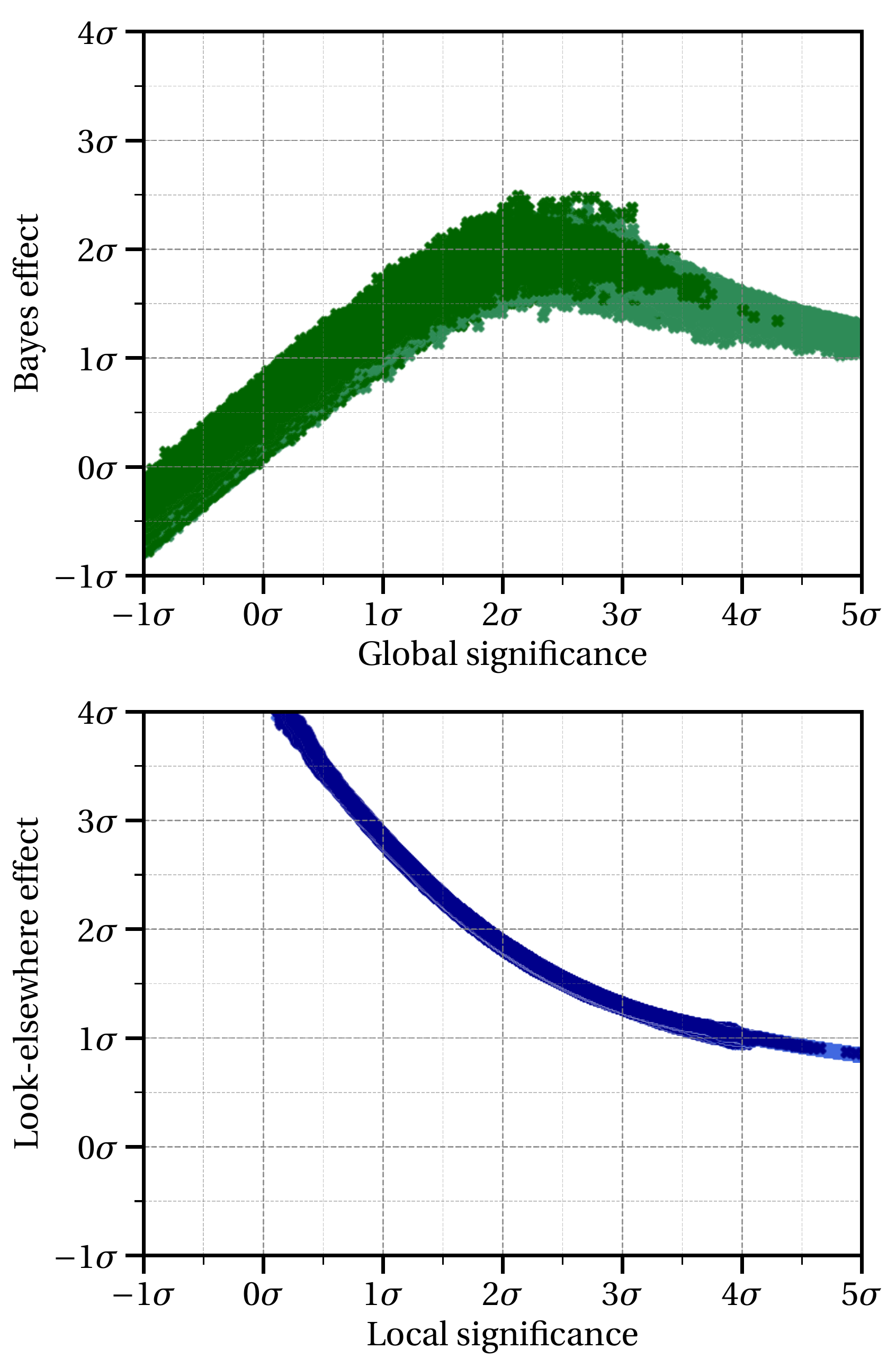}
    \caption{The reduction in the global significance by the Bayes effect (top panel) and the reduction in the local significance by the look-elsewhere effect (bottom panel). Results from data drawn from the background only model are shown in darker colors.}
    \label{fig:effect}
\end{figure}

We convert posteriors and \pvalues to \zvalues by the one-tailed rule in \refeq{eq:z_value}. In \reffig{fig:effect} (top panel) we show the difference between the global significance (i.e., the global \pvalue converted to a \zvalue) and the Bayesian significance (i.e., the posterior of the background model converted to a \zvalue) as a function of the global significance. We call this difference the Bayes effect, i.e.,
\begin{equation}\small
\text{Bayes effect} \equiv \text{Global significance} - \text{Bayesian significance}.
\end{equation}
This can be interpreted as a calibration of global significances. For global significances greater than about $1\sigma$ the Bayes effect is at least $1\sigma$ and peaks at over $2\sigma$ when the global significance is about $2\sigma$.
Thus the Bayes effect means that in the Bayesian framework global significances of about $2\sigma$ vanish and even global significances of $5\sigma$, the discovery threshold, reduce to about $4\sigma$. The Bayes effect appears as important as the look-elsewhere effect in resonance searches (bottom panel) which reduces local significances by a similar amount.\footnote{Our numerical results are consistent with the fact that the ratio of the global to local \pvalues\ --- the trials factor --- increases linearly with the local significance. The apparent contradiction with the fact that the change in significance decreases as the local significance increases stems from the non-trivial conversion to \zvalues.} We anticipated that the Bayes effect would be non-zero as we knew that the posterior and \pvalue would not be identical; however, we find the magnitude of the Bayes effect interesting, as it could change our interpretation of anomalies in resonance searches. There is spread in the Bayesian significance as it is a function of the entire dataset and not only the test-statistic. The Bayes effect occurs regardless of whether data were generated from the background (dark colors) or signal model (light colors) and at all luminosities under consideration.

As discussed in \refsec{sec:freq}, global \pvalues greater than $0.001$ were calculated with a bootstrap technique and those smaller than that were calculated with Gross-Vitells. A minor discrepancy between the techniques is visible at the boundary at $\pvalue = 0.001$ in \reffig{fig:difference} (green points) and at $\zvalue \approx 3$ in \reffig{fig:effect} (bottom panel). The discrepancy --- about $1\%$ in \zvalue\ --- originates from Monte-Carlo errors in the bootstrap technique and asymptotic approximations in Gross-Vitells.

Finally, it is tempting but risky to assume that the magnitude of the Bayes effect in this setting translates directly to other tests of the SM. Although a general result --- the Vovk-Selke bound on the Bayes factor\cite{vovk,doi:10.1198/000313001300339950} --- indicates that effects at least as great as $1\sigma$ could be common, its assumptions may be violated in tests of the SM, as they are in this one. Thus whilst we expect a similar effect in any test of the SM and there might be some useful rules of thumb, the actual result must be problem specific and requires a full calculation. In this regard, the Bayes effect is similar to the look-elsewhere effect.

\section{Sensitivity to our choices of prior}\label{sec:sensitivity}

\begin{figure*}[t]
    \centering
    \includegraphics[width=0.9\textwidth]{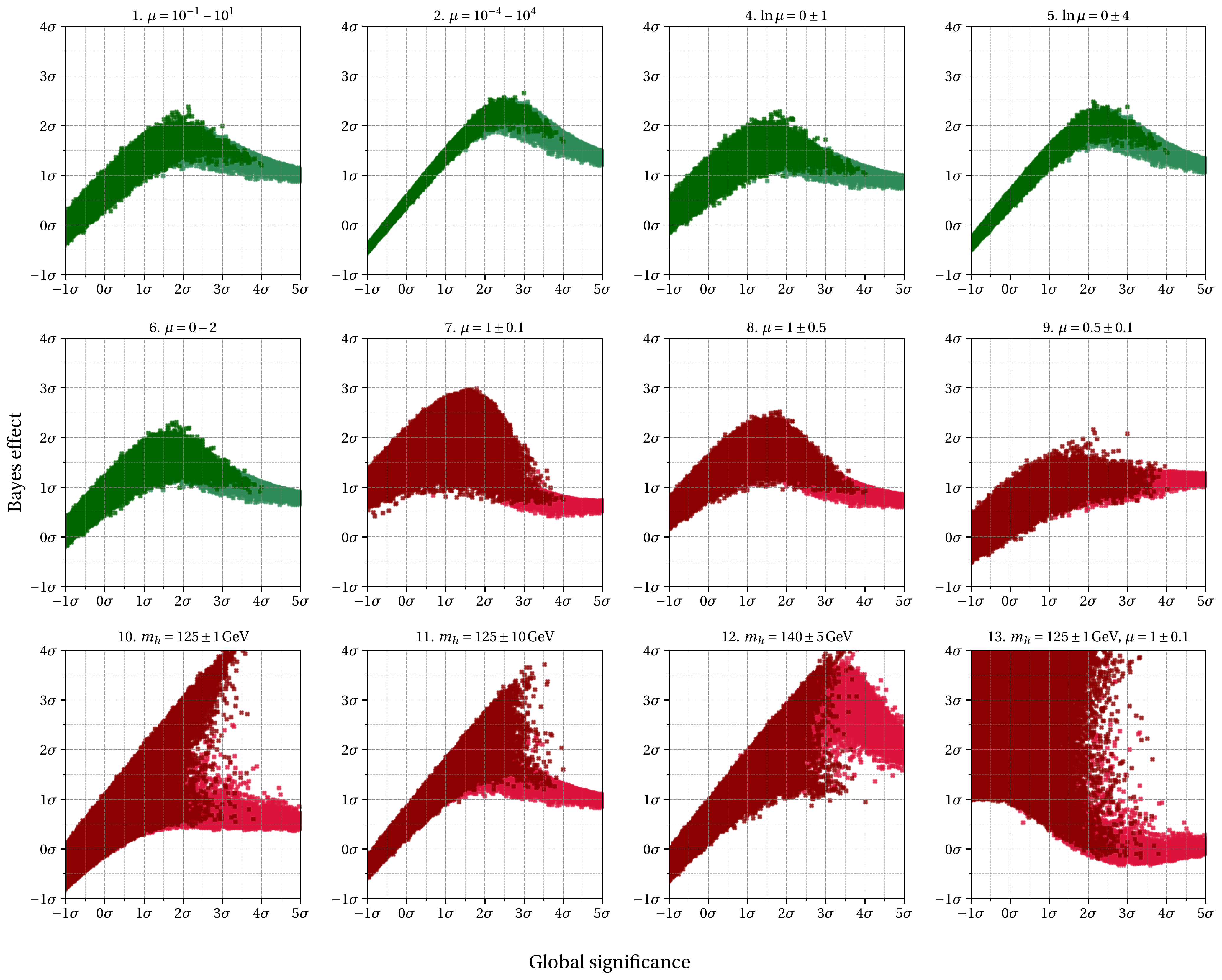}
    \caption{Reduction in the global significance by the Bayes effect for several choices of vague prior (green colors) and specific prior (red colors). We scatter data generated under the background model (dark) and the signal model (light) with $25\invfb$. The number of the plot corresponds to the numbered list of priors in \refsec{enu:priors}.}
    \label{fig:prior}
\end{figure*}

\begin{figure}[t]
    \centering
    \includegraphics[width=\figwidth]{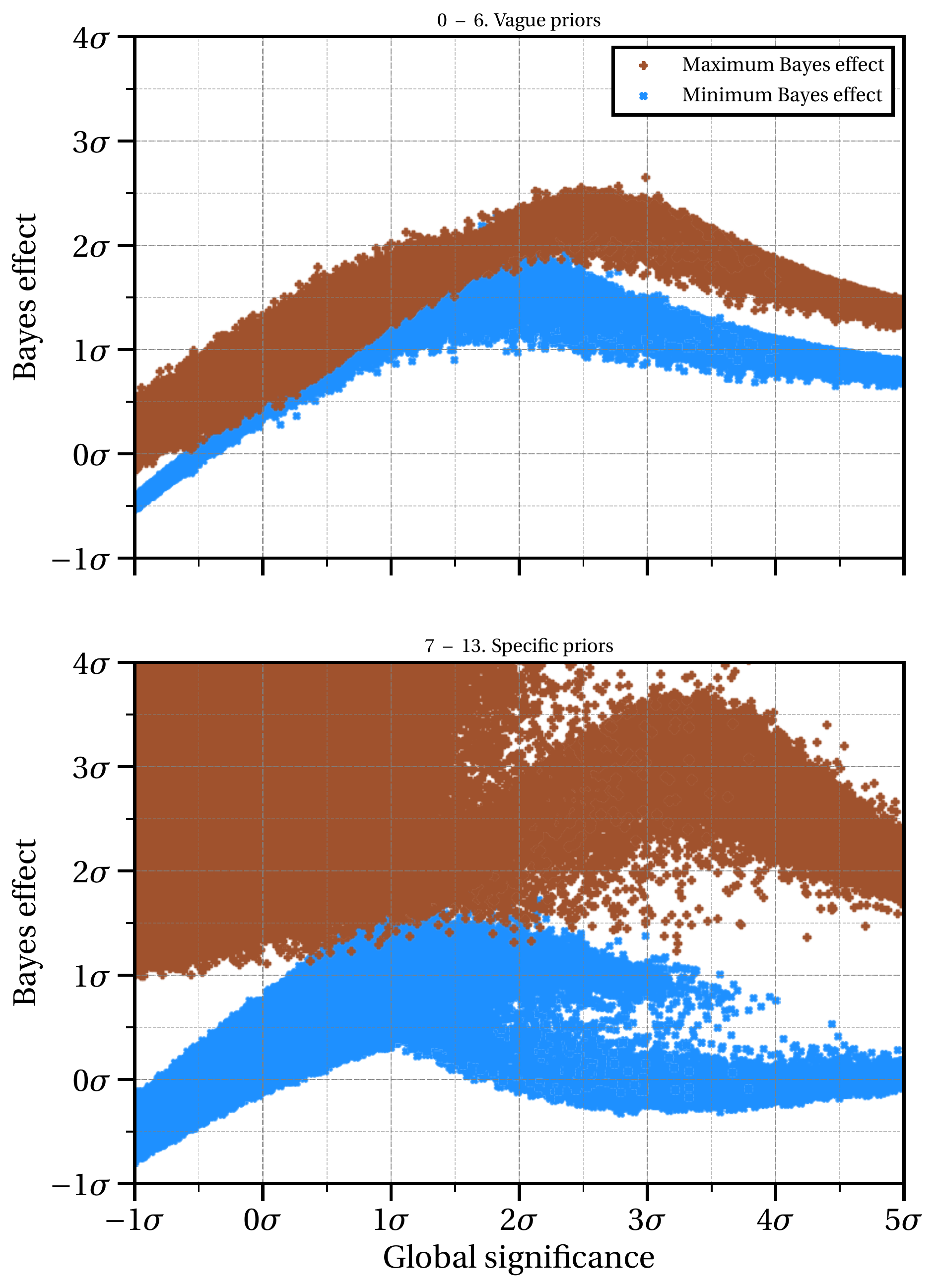}
    \caption{Summary of prior sensitivity of the Bayes effect: we show the maximum (brown) and minimum (blue) Bayes effects found from our \nvague vague priors (top, green panels of \reffig{fig:prior}) and our \nspecific specific priors (bottom, red panels of \reffig{fig:prior}).}
    \label{fig:sensitivity}
\end{figure}

The posterior of the background only model depends upon choices of prior for the Higgs mass and signal strength. To check whether the Bayes effect was an artifact of our particular choices, we repeated our calculations at $25\invfb$ with \naltprior alternative choices of prior:
\begin{enumerate}[label=\textrm{\arabic*.}, ref=\textrm{\arabic*}]\label{enu:priors}
\interlinepenalty=10000 % Penalize a page break inside an item
\setcounter{enumi}{-1} % Start at 0

\item\labelenu{item:default} Our default choice; the priors described in \refsec{sec:bayes}.

\item We halved the breadth of the prior for the logarithm of the signal strength so that $\log_{10}\mu$ spanned $-1$ to $1$.

\item\labelenu{item:broad} We doubled the breadth of the prior for the logarithm of the signal strength so that $\log_{10}\mu$ spanned $-4$ to $4$.

\item\labelenu{item:log_mass} We picked a logarithmic prior for the Higgs mass.

\addtocounter{enumi}{1} % Move onwards manually

\item[4.\todash5.]\labelenu{item:narrow} We picked a Gaussian prior for the order of magnitude of the signal strength, centered at $\ln \mu = 0$. We made two choices for the width, $\sigma_{\ln\mu} = 1$ and $\sigma_{\ln\mu} = 4$.

\addtocounter{enumi}{1} % We skip one item

\item\labelenu{item:flat} We picked a flat prior for the signal strength, $\mu = 0 \todash 2$.

\addtocounter{enumi}{1} % Move onwards manually

\item[7.\todash 9.] We picked a fat-tailed distribution, a Cauchy truncated to allow only positive values, for the signal strength. We centered it at $\mu=1$, representing prior knowledge that $\mu\approx1$ and made two choices for the width, $\gamma = 0.1$\labelenu{item:specific_mu} and $\gamma = 0.5$.
We furthermore picked one representing prior information that was faulty: $\mu = 0.5$ with a width $\gamma = 0.1$.

\addtocounter{enumi}{3} % We skip two items and move onwards manually

\item[10.\todash12.] We picked a Gaussian prior for the Higgs mass, centered at $m_h = 125\gev$. We made two choices for the width, $\sigma = 1\gev$\labelenu{item:specific_mh}
and $\sigma = 10\gev$. We again picked one representing information that was faulty: $m_h = 140\gev$ with a width $\sigma = 5\gev$.
\addtocounter{enumi}{2}% We skipped 11 and 12
\labelenu{item:wrong_mass}

\item\labelenu{item:knew_both} Finally, we considered a prior representing accurate, precise prior information about the mass and strength: Gaussians centered at $m_h = 125\gev$ and $\mu=1$ with widths $1\gev$ and $0.1$, respectively.
\end{enumerate}
The priors represented different possible states of knowledge about the Higgs mass and coupling strength; from vague knowledge, represented by broad priors, to specific information about their likely values, represented by distributions peaked at e.g., $\mu=1$ or $m_h = 125\gev$. If we specified the prior for only the Higgs mass or signal strength, we used the choice of prior introduced in \refsec{sec:bayes} for the other.

We show the resulting Bayes effects in \reffig{fig:prior}. We first consider vague priors (green colors). The Bayes effect shows mild dependence on the prior. For substantial global significances the effect is stronger for broader priors, e.g., the effect was biggest for a logarithmic prior on the signal strength between $10^{-4}$ and $10^4$ (\refprior{item:broad}) and weakest for a log-normal prior that favored a particular order of magnitude (\refprior{item:narrow}) and the flat prior (\refprior{item:flat}). This was anticipated as the evidence for the signal model could be diluted by picking a broader prior, enhancing the posterior of the background model. Global significances of about $1\todash4\sigma$, however, were always reduced by at least about $1\sigma$. As it was so similar to the result from \refprior{item:default}, we do not show the result for \refprior{item:log_mass}.

Second, we considered priors representing hypothetical specific prior knowledge about the Higgs mass or signal strength (red colors). In this case, the behavior depends strongly on whether data was generated under the signal or background model, since the latter may generate resonance-like features that do not agree with our prior knowledge for the mass and strength of the resonance. We find that for data generated under the signal model, specific correct knowledge reduces the Bayes effect, since it reduces the required tuning. For data generated under the background model, however, the Bayes effect could be amplified, as the posterior of the background is enhanced if the anomaly is not in accord with prior knowledge, regardless of its global significance. This resulted in substantial variation in the posterior associated with a particular global significance (see e.g., \refprior{item:specific_mu} and \refprior{item:specific_mh}).

If we had roughly known both the Higgs mass and signal strength in advance (\refprior{item:knew_both}), there would be only a minor Bayes effect under the signal model if the global significance was greater than about $4\sigma$. The Bayes effect could, however, reduce global significances of about $2\sigma$ to nothing. Under the background model, the Bayes effect was potentially dramatic, reducing global significances of less than about $3\sigma$ to evidence against the signal model, though showed substantial variance.

Finally, we consider specific but faulty prior information. Under the signal model this enhances the posterior of the background and thus the Bayes effect, since the prior information conflicts with the observed resonance-like feature. For the prior centered at $m_h=140\pm5\gev$ (\refprior{item:wrong_mass}), the Bayes effect becomes dramatic, reducing global significances of $5\sigma$ to about $3\sigma$.

We summarize the prior sensitivity of the Bayes effect in \reffig{fig:sensitivity}. For the vague priors (top panel), the minimum and maximum effects are typically separated by less than about $1\sigma$. For moderate global significances, the Bayes effect always reduces the significance by at least about $0.5\sigma$. For the specific priors (bottom panel), the maximum effect is typically substantial, as specific information that conflicts with data drastically reduces the significance. We stress, however, that the minimum effect is typically greater than about $0\sigma$; negative Bayes effects that dramatically increase the significance are impossible. This is consistent with the fact that the minimum posterior achievable for any prior was only about a factor four smaller than the global \pvalue (\reffig{fig:difference}).

\section{Sensitivity to the integrated luminosity and the Jeffreys-Lindley paradox}\label{sec:jl}

\begin{figure}[t]
    \centering
    \includegraphics[width=\figwidth]{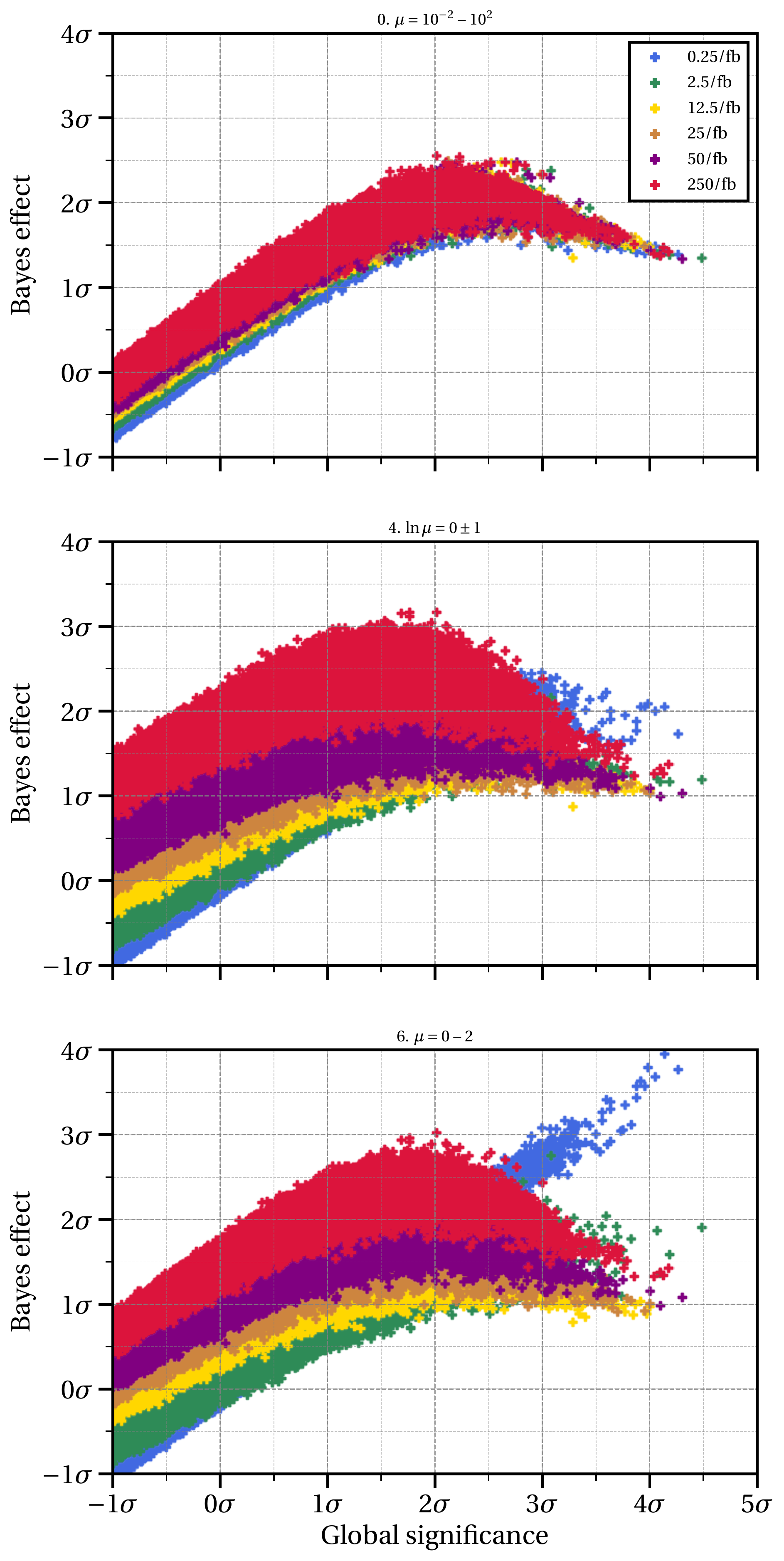}
    \caption{The Bayes effect at several integrated luminosities (indicated by different colors) for a logarithmic (top), log-normal (middle) and flat (bottom) prior for the signal strength. The number of the plot corresponds to the numbered list of priors in \refsec{enu:priors}. The data were generated under the background only model.}
    \label{fig:jl}
\end{figure}

To investigate the paradox discussed in \refsec{sec:intro_jl}, we repeated our calculations at $0.25\invfb$, $2.5\invfb$, $12.5\invfb$, $25\invfb$, $50\invfb$ and $250\invfb$ for three of our choices of prior for the signal strength: our default choice (\refprior{item:default}), a logarithmic prior that favors no particular scale; a log-normal prior (\refprior{item:narrow}), which favors $\ln\mu\simeq 0\pm1$; and a flat prior between $0$ and $2$ (\refprior{item:flat}). In \reffig{fig:jl} we show the Bayes effect as we increase integrated luminosity for these three choices. We only consider data generated under the background only; we cannot easily consider fixed significance under the signal model as the significance increases as we increase luminosity. The logarithmic prior results in mild dependence on the luminosity. The Bayes effect, however, shows strong dependence on the luminosity for a log-normal and flat prior. 

We can understand the absence of a paradox for a logarithmic prior by reconsidering the canonical example. In \refeq{eq:bf_jl_approx}, the Bayes factor's explicit dependence on $n$ cancels for a logarithmic prior, $\prior{\mu} \propto 1/\mu$, since it would result in $\prior{\mu = \bar x} \propto \sqrt{n}$. The fact that the Bayes factor no longer explicitly depends on $n$ is not surprising as the logarithmic prior is invariant under changes of scale, and by considering more data at fixed chi-squared we are probing smaller scales. This result in fact holds exactly in \refeq{eq:bf_jl} as for a logarithmic prior that equation is independent of $n$ at fixed chi-squared, which may be shown by rescaling the integration variable $\mu \to \mu / \sqrt{n}$ and fixing the chi-squared by setting $\bar x = \chi \sigma / \sqrt{n}$.

There is though a residual dependence on scale as to ensure that it was proper the logarithmic prior spanned only $10^{-2}$ to $10^2$. The limits of integration in \refeq{eq:bf_jl} thus reintroduce a dependence on scale, especially at small global significances when the likelihood isn't sharply peaked. Indeed, at small global significances the Bayes effect mildly increases with increasing luminosity, as the data disfavor an increasingly greater fraction of the range $10^{-2}$ to $10^2$. This reduces the evidence for the signal model and thus increases the Bayes effect.

The presence of the paradox for a log-normal prior stems from the fact that it favors a particular magnitude, $\ln \mu \simeq 0$. The Bayes effect thus depends on the signal strength preferred by the global significance and luminosity --- at fixed significance, the preferred signal strength decreases with luminosity; at fixed luminosity, it increases with significance. Surprisingly though, the Bayes effect at $3\sigma$ appears as big at $0.25\invfb$ as at $250\invfb$. This occurs because in the former case the preferred magnitude is greater than one, whereas in the latter case it is less than one. Thus in both cases the preferred signal strength conflicts with the prior resulting in appreciable Bayes effects. 

Relative to the logarithmic prior, the flat prior favors greater orders of magnitude for the signal strength, $\prior{\ln \mu} \propto \mu$. The presence of the paradox, therefore, is not surprising, as at fixed chi-squared the data favors increasingly smaller scales as the luminosity increases. With only $0.25\invfb$ and moderate global significances, the data favor signal strengths greater than 2. The prior, however, only supports signal strengths less than 2, resulting in substantial Bayes effects. 

In summary, owing to the Jeffreys-Lindley paradox, the Bayes effect depends on an interplay between the luminosity and the prior. Since we are warning about the presence of a substantial Bayes effect in this context, an effect that increases it and makes it more unpredictable does not alter our conclusions.

\section{Conclusions}\label{sec:conclusions}

We compared Bayesian and frequentist approaches to resonance searches using toy experiments based on a Higgs search in the diphoton channel. We first found two pedagogical results. First, with increasing luminosity, we showed that the Bayesian posterior always converges to overwhelming favor the correct model, whereas the \pvalue makes a random walk if the null is true, and otherwise converges to $0$. Second, we showed frequentist properties of the Bayesian posterior, helping shed light on Bayesian results for those with frequentist mindsets. We found that the type-1 error rate from a threshold on the posterior, whilst not under direct control, was significantly less than the threshold, i.e., if we reject the background only model when the posterior is less than $0.05$, the type-1 error rate is significantly less than $0.05$ in this setting. The ROC curves using the posterior and log likelihood ratio as test-statistics were extremely similar.

Our final finding, however, was striking; in our canonical resonance search toy problem, global significances were typically greater than those from our Bayesian framework by about $1\sigma$ to $2\sigma$, e.g., anomalies that appear at global significance of $3\sigma$ correspond to only about $1\sigma$ evidence in the Bayesian framework. This effect --- which we call the Bayes effect --- was robust with respect to \nvague vague choices of priors for the mass and signal strength, though slightly decreased when the breadth of the prior was reduced. For our \nspecific choices representing specific prior knowledge about the mass or strength, we found that the effect persisted, though its magnitude strongly depended on whether the prior information conflicted with the observed resonance-like feature. For a logarithmic prior the effect depended extremely mildly upon the integrated luminosity, though for other priors a Jeffreys-Lindley paradox was observed, as anticipated.

We showed that when we discarded information, and worked assuming that we only knew that the test-statistic was at least as big as a threshold, we found similar results from Bayesian and frequentist methods. This suggests that the Bayes effect originates from the fact that \pvalues use only the fact that the test-statistic was at least as big as that observed, whereas the posterior uses the exact observed data.
% The Bayes effect should dramatically change our interpretation of global significances as evidence for new physics in resonance searches.

% \begin{acknowledgments}
% I would like to thank for helpful comments and advice about my manuscript.
% \end{acknowledgments}

\bibliography{references}
\end{document}